\documentclass[12pt]{article}

\newcommand{\sect}[1]{ \section{#1} }

\newcommand{\ve}{\left( \begin{array}{r}}
\newcommand{\ev}{\end{array} \right)}
\newcommand{\ar}{\left( \begin{array}{rr}}
\newcommand{\ra}{\end{array} \right)}
\newcommand{\arr}{\left( \begin{array}{rrrr}}
\newcommand{\arrr}{\left( \begin{array}{rrrrrr}}

\newcommand{\eqr}{\begin{eqnarray}}
\newcommand{\rqe}{\end{eqnarray}}
\newcommand{\eq}{\begin{equation}}
\newcommand{\qe}{\end{equation}}

\def\KK{{\rm I\kern -.2em  K}}
\def\NN{{\rm I\kern -.16em N}}
\def\RR{{\rm I\kern -.2em  R}}

\def\ZZZ{{\small{\rm Z}\kern -.5em Z}}
\def\QQ{{\rm \kern .25em
             \vrule height1.4ex depth-.12ex width.06em\kern-.31em Q}}
\def\CC{{\rm \kern .25em
             \vrule height1.4ex depth-.12ex width.06em\kern-.31em C}}

\title{Warped AdS near-horizon geometry of completely localized 
intersections of
 M5-branes}
\author{Ansar Fayyazuddin$^1$\footnote{email: ansar@physto.se}
\quad and \quad
Douglas J. Smith$^2$\footnote{email: Douglas.Smith@durham.ac.uk} }

\begin{document}

\maketitle

\begin{center}

{\em
$^1$Department of Physics\\
Stockholm University\\
Box 6730\\
S-113 85 Stockholm\\
Sweden \\
\vspace{0.5cm}
$^2$Department of Mathematical Sciences \\
University of Durham \\
Durham \\
DH1 3LE \\
UK
}

\end{center}

\vspace{1.4cm}

\begin{abstract}
We study a Hanany-Witten set-up relevant to ${\cal N}=2$
superconformal field theories.
We find the exact near-horizon solution for this 11-dimensional 
system which involves intersecting M5-branes.  The metric describes
a warped product of $AdS_5$ with a
manifold with $\mathrm{SU}(2)\times \mathrm{U}(1)$ isometry.

\end{abstract}

\vspace{-20cm}
\begin{flushright}
USITP-00-08\\
DTP/00/45\\
hep-th/0006060\\
\end{flushright}

\thispagestyle{empty}

\newpage

\setcounter{page}{1}

\sect{Introduction}
Intersecting branes are ubiquitous in string theory.  Although much
effort has been put into finding supergravity solutions of these
systems, only solutions with branes localized in the overall 
transverse \cite{gaunt}
dimensions and those with at most one set of completely localized 
branes are 
known \cite{Itzhaki:1998uz}-\cite{loc6}.  The cases where all branes are 
localized 
have so far eluded solution.  

In the past year some progress has been made in finding
solutions of partially localized intersecting branes. Some
of these results can be found in \cite{Itzhaki:1998uz}-\cite{loc6}.
An interesting development has been to 
interpret brane delocalization physically \cite{pema}.
In this approach the delocalization seen in the supergravity
solution is interpreted via the AdS/CFT correspondence of Maldacena
\cite{Maldacena, AdS_refs}
as a Coleman-Mermin-Wagner theorem in the field theory.
These results do not directly apply to the case we study and
we will see that our near-horizon geometry 
describes completely localized branes.

In this paper we report an exact solution of 11-dimensional 
supergravity for a system of intersecting M5-branes in the near-horizon limit.  
The particular system we study is the supergravity dual of ${\cal N}=2$
superconformal field theory with gauge group $\mathrm{SU}(N)$ and
$N_{f}= 2N$ fundamental flavors.  This paper is a continuation
of our work \cite{danda} in which we solved the supersymmetry 
preservation conditions for the system.  The full solution requires
solving for a K{\" a}hler metric satisfying a non-linear partial 
differential equation in 7 variables!  In our previous paper 
\cite{danda} we solved this equation in an approximation where
one set of branes were localized while the second set were smeared out
over the worldvolume directions of the first set (these partially
localized solutions were also found independently in 
\cite{loc3,loc3.5}.)
This equation was studied in \cite{loc4} to yield
an iterative expansion around the asymptotically flat
region.  This is the opposite limit to the one we pursue here.
In the present paper we solve this differential equation exactly in
the near horizon limit which is relevant to the AdS/CFT duality
\cite{Maldacena}.

The paper is structured as follows.  We start with a brief description
of the system under study.  We then take a scaling limit
where the Planck scale is taken to infinity while keeping field 
theory 
quantities fixed.  Finally, we solve for the metric in this 
``near-horizon''
limit, finding a warped AdS geometry. Warped AdS metrics have recently been
discussed in the context of the AdS/CFT correspondence and 
semi-localized intersecting branes
\cite{WAdS1, WAdS2} for brane configurations similar to ours. 
We conclude with some comments. 

\section{The system}
In this section we set up the problem and summarize some results from 
\cite{danda} which we will need.

One way of studying ${\cal N}=2$ gauge theories is to generalize the 
Hanany-Witten \cite{HW} set-up to a system relevant to 
four dimensional gauge theories \cite{witn2}.  
The idea \cite{witn2} is to suspend D4-branes between a pair of 
NS5-branes
which are separated by a finite coordinate distance $L$.  
The gauge theory living on the D4-branes will be, in the infrared,
a four dimensional Yang-Mills theory with ${\cal N}=2$ supersymmetry and
gauge group $\mathrm{SU}(N)$.  There are many ways of introducing fundamental 
matter, but the easiest method is to
introduce semi-infinite D4-branes on either side of the
NS5-branes. The gauge D4-branes detect the semi-infinite D4-branes
through strings which have ends on both types of D4-branes.  These
strings carry Chan-Paton factors with respect to the gauge groups
of both types of D4-branes. From the gauge theory point of view
these represent fundamental matter transforming in the fundamental
representation of (a subgroup of) the flavor group.  

Witten pointed out that this system can be lifted to M-theory where
this web of D4-branes and NS5-branes can be viewed as a single 
M5-brane
wrapping a non-compact Riemann surface which coincides with the
Seiberg-Witten Riemann surface.  The same picture was derived in a 
different way in \cite{vafa}.  

In the remainder of this paper we will study a configuration of 
branes 
consisting of a set of coincident infinite D4-branes intersecting a 
pair of 
separated NS5-branes.  
This configuration can be viewed as one particular realization of 
$\mathrm{SU}(N)$
gauge theory with $N_{f}=2N$ according to the recipe described 
above.   Our set-up can be arrived at from any generic 
Hanany-Witten configuration describing this field theory by moving the
D6-branes (on which the semi-infinite D4-branes end `at infinity')
through the NS5-branes so that we have an equal number of semi-infinite
D4-branes on both sides. We 
have also tuned the moduli so that all the D4-branes are coincident and
collinear.  In the gauge theory this corresponds to both tuning the
bare masses of the fundamental matter to zero and sitting at the
origin of the Coulomb branch where the gauge group is enhanced to the
full $\mathrm{SU}(N)$.  When viewed from the point of view of 
M-theory this looks simply like a system of intersecting M5-branes.

It is convenient to pick a coordinate system such that the $N$ 
D4-branes
have world-volume directions along $x^{0}, x^{1}, x^{2}, x^{3}, x^{6}$
while the NS5-branes have world-volume directions along
$x^{0}, x^{1}, x^{2}, x^{3}, x^{4}, x^{5}$.  The two sets of branes 
then
intersect along $x^0, x^{1}, x^{2}, x^{3}$.  The positions of the 
NS5-branes
are $x^{6}=\pm L/2$.  This configuration of branes can be lifted to
M-theory with two sets of M5-branes intersecting along 
$x^0, x^{1}, x^{2}, x^{3}$.  Let us denote one set as M5(1) branes, they have
world-volume directions along $x^{0}, x^{1}, x^{2}, x^{3}, x^{6}, 
x^{7}$
and the other two M5-branes as M5(2), they have world volume directions
along $x^{0}, x^{1}, x^{2}, x^{3}, x^{4}, x^{5}$.  The M5(1) branes 
descend
to D4-branes when $x^7$ is compactified to give type IIA string 
theory
while the M5(2) branes are localized in the 
compactified $x^7$ direction and become NS5-branes.  
It is convenient to define a complex structure in the subspace
$x^{4}, x^{5}, x^{6}, x^{7}$ as follows:
\eqr
v \equiv z^1 & = &x^4 + ix^5\\
s \equiv z^2 & = &x^6 + ix^7.
\rqe
We also take $x^7$ to be a compact direction with radius $R$.  

In \cite{danda} we solved the supersymmetry variation equations
for M5-brane configurations which preserve at least 8 real 
supersymmetries.
The general solution is given by the metric:
\begin{equation}
ds^2 = g^{-\frac{1}{3}}dx_{3+1}^2 +
        g^{-\frac{1}{3}}g_{m\overline{n}}dz^mdz^{\overline{n}} +
        g^{\frac{2}{3}}\delta_{\alpha \beta}dx^{\alpha}dx^{\beta},
\end{equation}
and the 4-form field strength:
\eqr
F_{m \overline{n} \alpha \beta} & = &
        \frac{i}{4} \epsilon_{\alpha \beta \gamma} \partial_{\gamma} 
		g_{m \overline{n}} \label{F_g_start} \\
F_{m89(10)} & = & -\frac{i}{2} \partial_m g \\
F_{\overline{m}89(10)} & = & \frac{i}{2} \partial_{\overline{m}} g.
\label{F_g_end}
\rqe
The Greek indices run over the overall transverse coordinates 
$x^{8},x^{9},x^{10}$.  Both the metric and 4-form are expressed
in terms of the K{\" a}hler metric $g_{m\overline{n}}$.  The source
equations for the 4-form $F$ force $g_{m\overline{n}}$ to satisfy
the non-linear partial differential equations:
\begin{equation}
\partial_{\gamma}\partial_{\gamma} g_{m\overline{n}} +
4 \partial_m\partial_{\overline{n}} g = J_{m \overline{n}}
\label{source}
\end{equation}
where $J$ is the source specifying the positions of the M5-branes. 
The quantity $g$ appearing in the above equations 
is the square root of the determinant of the K{\"a}hler metric: 
$g=g_{v \overline{v}}g_{s \overline{s}}-g_{v \overline{s}}g_{s 
\overline{v}}$.

For the particular configuration that we will be studying the 
source equations are:
\begin{eqnarray}
\nabla^2 g_{s\overline{s}} + 4\partial_{s}\partial_{\overline{s}}g
& = & -8 \pi^3l_{p}^{3}\delta^{(3)}(r)(\delta^{(2)}(s-L/2)
+\delta^{(2)}(s+L/2)) \nonumber \\
\nabla^2 g_{v\overline{v}} + 4\partial_{v}\partial_{\overline{v}}g
& = & -N8\pi^3l_{p}^{3}\delta^{(3)}(r)\delta^{(2)}(v) \nonumber \\
\nabla^2 g_{v\overline{s}} + 4\partial_{v}\partial_{\overline{s}}g
& = & 0
\end{eqnarray}
where $\nabla^2$ is the flat Laplacian in the overall transverse space.

To summarize, a given M5-brane configuration determines a source $J$ 
in (\ref{source}).  Solving this source equation for the 
metric $g_{m\overline{n}}$ then determines all other quantities
in the supergravity solution.

\sect{Field theory (``near-horizon'') limit}
Maldacena proposed a certain scaling limit of string theory quantities
\cite{malstrings, Maldacena} to isolate the world-volume gauge theory 
from bulk interactions.  
The idea is simply to take a limit in which the Planck length goes
to zero while keeping field theory quantities fixed.  In
\cite{danda} we pointed out the relevant scalings of supergravity
variables in type IIA theory.  We can express these scalings in
M-theory units by defining $w, t$ and $y$ as follows:

\eqr
w & = & \frac{v}{\alpha'} = \frac{vR}{l_{p}^{3}} \nonumber\\
t^2 & = & \frac{r}{g_s\alpha'^{\frac{3}{2}}} = \frac{r}{l_{p}^{3}} \\
y & = & \frac{s}{R}\nonumber.
\rqe
Note that $w$ and $y$ are complex variables while $t$ is a real
variable.  
The field theory limit is one in which we keep $w,y,t$
fixed while taking $l_{p}$ to zero.  
We take the metric to be:
\begin{equation}
\frac{1}{l_{p}^2}ds^2 = g^{-\frac{1}{3}}\eta_{\mu 
\nu}dx^{\mu}dx^{\nu} +
        g^{-\frac{1}{3}}g_{m\overline{n}}dz^mdz^{\overline{n}} +
        g^{\frac{2}{3}}(4t^2 dt^2 + t^4 d\Omega_{2}^2)
\label{soln_11}
\end{equation}
Where now $m,n$ run over $y,w$, 
$g=g_{w \overline{w}}g_{y \overline{y}}-g_{w \overline{y}}g_{y 
\overline{w}}$ and $d\Omega_{2}^2$ is the metric on the round unit
2-sphere.
The source equations become:
\begin{eqnarray}
\frac{1}{4t^5}\partial_{t}(t^3\partial_{t}) g_{y\overline{y}} + 
4\partial_{y}\partial_{\overline{y}}g
& = & -\pi^2\frac{\delta (t)}{t^5} \left( \delta^{(2)}
(y-\frac{1}{2g_{YM}^2})
+\delta^{(2)}(y+\frac{1}{2g_{YM}^2}) \right) \nonumber \\
\frac{1}{4t^5}\partial_{t}(t^3\partial_{t}) g_{w\overline{w}} + 
4\partial_{w}\partial_{\overline{w}}g
& = & -N\pi^2\frac{\delta (t)}{t^5}\delta^{(2)}(w) 
\nonumber \\
\frac{1}{4t^5}\partial_{t}(t^3\partial_{t}) g_{w\overline{y}} + 
4\partial_{w}\partial_{\overline{y}}g
& = & 0
\end{eqnarray}
where $g^{2}_{YM}= R/L$ is the Yang-Mills coupling constant in the 
field theory. We have
also assumed that the metric depends only on the ``radial'' coordinate
$t$ and not the ``angular'' variables in the overall transverse 
directions. 
This is consistent with the requirement of having an $\mathrm{SU}(2)$ isometry 
corresponding to the field theory R-symmetry.  

We end this section with some comments.
The source equations now have no powers of the Planck length $l_{p}$,
they are expressed in terms of quantities which have a field theory 
interpretation.  This will allow us to express the metric in terms
of only these rescaled variables with no further dependence on 
$l_{p}$ aside from the overall multiplicative factor.  Secondly,
we would like to point out that while the initial set up treated
the M5(1) and M5(2) branes on an equal footing, the scaling limit
we take breaks that symmetry.  This is clear from the type IIA picture
since there the D4-branes play a distinguished role in that the field
theory of interest lives on their world-volume.  

\section{Near-horizon geometry of intersecting M5-branes}
Since we are looking for a supergravity dual of a four-dimensional conformal
field theory, we expect a solution where the metric contains an $AdS_5$ 
factor.
The most general metric of this type is:
\begin{equation}
\frac{1}{l_P^2}ds^2 =
	\Omega^2 \left ( u^2 dx_{3+1}^2 + \frac{1}{u^2} du^2 \right ) + 
ds_6^2
\end{equation}
where the metric for the six-dimensional transverse space and 
$\Omega$ are
independent of the $AdS_5$ coordinates.  

It is convenient to choose variables where there is only one 
dimensionful
variable. We choose to do this by defining $\rho$ with dimensions of 
mass and
dimensionless angular variables $\theta$ and $\phi$ by:
\begin{eqnarray}
t & = & \rho \cos\theta \\
w & = & \rho \sin\theta e^{i\phi}
\end{eqnarray}

Now we see on dimensional grounds that $u$ must be related to $\rho$ 
by
$u = \rho \alpha$ where $\alpha$ is some function of the dimensionless
variables $\theta$, $\phi$, $y$ and $\bar{y}$. By substituting this 
expression
for $u$ into the above metric, we can compare the metric components 
with
those in the known form of the solution, eq.~(\ref{soln_11}). In particular, by examining 
the
factor multiplying $dx_{3+1}^2$ and the metric components $g_{\rho 
\rho}$,
$g_{\rho y}$, $g_{\rho \theta}$ and $g_{\rho \phi}$ we find:
\begin{eqnarray}
g & = & \frac{1}{\Omega^6 \alpha^6 \rho^6} \nonumber \\
g_{w \bar{w}} & = & \frac{\Omega^6 \alpha^4 - 4 \cos^4\theta}{\rho^4 
\Omega^6 \alpha^6 \sin^2\theta} \nonumber\\
g_{y \bar{w}} & = &
        \frac{2 e^{i\phi} \partial_y \alpha}{\rho^3 \alpha^3 
\sin\theta} \label{ads}\\
\partial_{\theta} \alpha & = & \frac{(\Omega^6 \alpha^4 - 4 
\cos^2\theta)\cos\theta}{\Omega^6 \alpha^3 \sin\theta} \nonumber\\
\partial_{\phi} \alpha & = & 0\nonumber
\end{eqnarray}
Since we are 
looking at ${\cal N}=2$ superconformal field theories we would like to 
preserve a $\mathrm{SU}(2)\times \mathrm{U}(1)$ isometry.  This we have incorporated
in the above ansatz by requiring that the metric preserve a $\mathrm{U}(1)$ 
which rotates $w$ by a phase and the $\mathrm{SU}(2)$ symmetry of
the transverse 2-sphere. In fact, these symmetries are consequences of our
required form of the metric. For example we see that $\alpha$ is independent of
$\phi$ and the equation for $\partial_{\theta}\alpha$ shows that $\Omega$ must
also be independent of $\phi$.

If we assume, for the moment, that $\Omega$ is constant\footnote{As
we will see below the metric has a warped AdS structure
and so our assumption that $\Omega$ is constant is incorrect.  
Nevertheless, for the purposes of solving the equations we find it
easier to begin with an incorrect assumption which can be easily modified
to yield the correct metric than to solve the equations directly.},
say $\Omega_0$, we can solve for the 
$\theta$-dependence
of $\alpha$ (which we denote by $\alpha_0$) in terms of an arbitrary function $A(y, \bar{y})$:
\begin{equation}
\Omega_0^6 \alpha_0^4 = 4\cos^4\theta + 4A(y, \bar{y})\sin^4\theta
\end{equation}
We can then write the above equations as:
\begin{eqnarray}
g & = & \frac{\Omega_0^3}{8 \rho^6 \beta^3} \label{g}\\
g_{w \bar{w}} & = & \frac{\Omega_0^3 A \sin^2\theta}{2 \rho^4 \beta^3} 
\\
g_{y \bar{w}} & = &
     \frac{e^{i\phi} \Omega_0^3 \sin^3\theta \partial_y A}{4 \rho^3 
\beta^3}
\end{eqnarray}
where we have defined:
\begin{equation}
\beta = \left ( \cos^4\theta + A\sin^4\theta \right ) ^{\frac{1}{2}}
\end{equation}

The metric component $g_{y \bar{y}}$ can be determined from the
determinant $g$ and the other components of the metric given in the 
above equations.  However, the metric determined in this way fails to be
K\"ahler \footnote{Hence the assumption of constant $\Omega$ is
incorrect.}. It is easier instead to determine $g_{y\bar{y}}$ by requiring the
metric to be K\"ahler. The K\"ahler condition is satisfied if:
\eq
g_{y\bar y} = \frac{\Omega_0^3 \sin^4\theta |\partial_y A|^2}{8 \rho^2 A \beta^3}
\qe
provided that $A = |F(y)|^2$, where $F$ is a holomorphic function of
$y$.  One can easily check that the source equations are satisfied
everywhere away from the support of the delta functions\footnote{The 
normalizations and precise form of $F$ relevant to the delta 
function sources are determined below.}.  The metric as it stands now
has a vanishing determinant so it is not a valid solution.  However,
it is easy to see that the metric can be modified in such a way as
to get the correct determinant while continuing to satisfy the
source equations.  The idea is simply to add to all the metric 
components additional terms which are themselves K\"ahler (so as
not to destroy the K\"ahler properties of our initial ansatz), such 
that the determinant is correctly reproduced.  The source equations
will continue to be satisfied provided these additional terms do not
depend on $t$. From these simple requirements one determines the 
solution:
\begin{eqnarray}
g_{w \bar{w}} & = & \frac{\Omega_0^3 A \sin^2\theta}{2 \rho^4 \beta^3} +
			\frac{|f|^2}{\rho^4 \sin^4\theta} \nonumber\\
g_{y \bar{y}} & = &
	\frac{\Omega_0^3 \sin^4\theta |\partial_y A|^2}{8 \rho^2 A \beta^3} + 
        \frac{|\partial_y f|^2}{\rho^2 \sin^2\theta} \label{metric}\\
g_{y \bar{w}} & = &
        \frac{e^{i\phi} \Omega_0^3 \sin^3\theta \partial_y A}{4 \rho^3 
\beta^3} -
        \frac{e^{i\phi} \bar{f} \partial_y f}{\rho^3 \sin^3\theta} 
        \nonumber
\end{eqnarray}
Where $f(y)$ is a holomorphic function determined from the requirement 
that the determinant has the form (\ref{g}).  This requirement can
be stated succinctly as a differential equation:
\begin{equation}
\partial_y(f^2F) = f
\end{equation}
The general solution of this equation is:
\begin{equation}
f(y) = \frac{\int_a^y F^{-\frac{1}{2}}(z) dz}{2 F^{\frac{1}{2}}(y)}
\label{soln_f}
\end{equation}
where $a$ is an arbitrary constant of integration.

Note that the additional terms in the metric (involving $f$) are 
independent
of $t$ ($r$ in the original coordinates) while the other terms in
$g_{m \bar{n}}$ can be expressed in the form $g \partial_m 
\partial_{\bar{n}}
(|Fw^2|^2)$. So the source equations can be conveniently written as:
\begin{equation}
\frac{1}{4t^5} \partial_t (t^3 \partial_t g) \partial_m 
\partial_{\bar{n}}
	(|Fw^2|^2) + 4 \partial_m \partial_{\bar{n}} g = J_{m \bar{n}}
\end{equation}
It is now a straightforward calculation to check that these equations 
are
satisfied with the source terms:
\begin{eqnarray}
J_{w \bar{w}} & = & -\pi^2 N \frac{\delta(t)}{t^5} 
\delta^{(2)}(w) \\
J_{y \bar{y}} & = & -\pi^2 \frac{\delta(t)}{t^5} \left (
\delta^{(2)}(y - \frac{1}{2g_{YM}^2}) + \delta^{(2)}(y + 
\frac{1}{2g_{YM}^2})
	\right ) \\
J_{y \bar{w}} & = & 0.
\end{eqnarray}
The source equations determine $F$ (and consequently also $f$) since 
the
M5(2) branes are localized at the zeroes of $F$, as well as fixing the
constant:
\begin{equation}
\Omega_0^3  =  4 \pi N.
\end{equation}
We 
will now give the explicit form of the metric, exhibiting the warped product
structure before solving for $F$. We will then consider the form of $F$ in the
large radius limit before solving it in 
the general case.

\subsection{Warped anti-de Sitter structure of the metric}
According to Maldacena's conjecture \cite{Maldacena} conformal field
theories have anti-de Sitter supergravity duals.  Our metric is not
a product manifold of anti-de Sitter space with a transverse manifold, 
but as mentioned earlier in this section, the metric can be written 
as a warped product consistent with Maldacena's conjecture.
To see this warped product structure, one simply returns 
to equations (\ref{ads}) and
solves for $\alpha$ and $\Omega$ using the explicit metric
appearing in (\ref{metric}). This yields:
\eqr
\alpha^{-2} & = &
\frac{2 \pi N}{(\cos^4\theta + |F|^2\sin^4\theta)^{\frac{1}{2}}}
+\frac{|f|^2}{\sin^2\theta} \nonumber \\
(\Omega\alpha)^{-6} & = & \frac{\pi N}{2}
\frac{1}{(\cos^4\theta + |F|^2\sin^4\theta)^{\frac{3}{2}}}.
\rqe
These expressions are consistent with all the metric components derived and it
can easily be checked that $\partial_{\theta}\alpha$ has the correct form
required by eq.~(\ref{ads}).
The metric, therefore, can be written as a warped product
of AdS space with a transverse manifold.  The metric, while messy,
can be written relatively concisely if one expresses it in terms
of $\alpha$ and $\Omega$:
\eqr
\frac{1}{l_P^2}ds^2 & = & \Omega^2 (u^2\eta_{\mu\nu}dx^{\mu}dx^\nu +
\frac{du^2}{u^2}) + \frac{4\cos^2\theta}{\Omega^4\alpha^4\sin^2\theta}
(1 - \frac{4\cos^4\theta}{\Omega^6\alpha^4})d\theta^2 \nonumber\\
& + &\frac{8\cos^3\theta}{\sin\theta\Omega^4\alpha^5}\partial_{y}\alpha
d\theta dy
+\frac{8\cos^3\theta}{\sin\theta\Omega^4\alpha^5}\partial_{\bar{y}}\alpha
d\theta d\bar{y} \nonumber \\
& + &\Omega^2(1-\frac{4\cos^4\theta}{\Omega^6\alpha^4})d\phi^2
-2i\Omega^2\frac{\partial_{y}\alpha}{\alpha}d\phi dy
+2i\Omega^2\frac{\partial_{\bar{y}}\alpha}{\alpha}d\phi d\bar{y}
\nonumber \\
& + &\frac{\Omega^2\alpha^2}{\Omega^6\alpha^4-4\cos^4\theta}
(\sin^2\theta + (2\Omega^6\alpha^4 + 8\cos^4\theta)
|\frac{\partial_{\bar{y}}\alpha}{\alpha^2}|^2)\|dy|^2 \nonumber \\
&-& \frac{\Omega^2}{\alpha^2}(\partial_{y}\alpha)^2dy^2
-\frac{\Omega^2}{\alpha^2}(\partial_{\bar{y}}\alpha)^2d\bar{y}^2
+\frac{\cos^4\theta}{\Omega^4\alpha^4}d\Omega_{2}^2
\rqe
Everything is now determined explicitly in terms of $F(y)$. We will now
consider the form of $F$ in various cases, including the simple generalisation
to conformal theories with gauge group $\mathrm{SU}(N)^n$.

\subsection{Large $R$ or M-theory limit}
In the limit that $R$, the radius of $x^7$, becomes infinite we can 
ignore the periodicity of $y$: $y\rightarrow y+i2\pi$.  Notice
that the field theory is not sensitive to the value of $R$ but
only to the ratio $R/L$ which determines the gauge coupling
constant.  We are thus simultaneously taking $L$ to infinity
while keeping $R/L$ fixed.  In this limit 
we can solve for $F$ taking into account the normalization of the 
sources:
\begin{equation}
F  =  \left( (y - \frac{1}{2g_{YM}^2})(y + 
\frac{1}{2g_{YM}^2}) \right)^{\frac{2}{N}}.
\end{equation}
Using this explicit expression we can calculate $f$:
\eq
f = \frac{1}{2}(4g_{YM}^4)^{\frac{1}{N}}y(y^2 
-\frac{1}{4g_{YM}^4})^{-\frac{1}{N}}
{\cal F}(\frac{1}{N}, \frac{1}{2};\frac{3}{2};4g_{YM}^4 y^2),
\qe
where ${\cal F}$ denotes the hypergeometric function.

It is easy to see how one can generalize this to an arbitrary number 
of M5(2) branes. If there are $n$ M5(2) branes located at $y=y_i$ then:
\eq
F = \prod_{i=1}^{n}(y-y_i)^{2/N}.
\qe
We can then determine, at least in principle, $f$ from this expression. The
dual conformal field theory will have a product gauge group 
$\mathrm{SU}(N)^{n-1}$
with the gauge coupling of the $i$'th factor being given by:
\begin{equation}
\frac{1}{g^2_{YM,i}} = y_{i+1} - y_{i}
\end{equation}

\subsection{Solution for arbitrary $R$}
As noted above the zeroes of $F$ determine the locations of the M5(2) 
branes.
From the previous section it is easy to see how to generalize to an 
arbitrary
radius of $x^7$ (i.e. when we take into account the periodicity of 
$y$).
For our sources with periodic $y$ the correct $F$ is:
\eq
F= \left( \sinh(y-\frac{1}{2g_{YM}^2}) \sinh(y+\frac{1}{2g_{YM}^2}) \right)^{2/N}.
\qe
In this case we have not been able to express $f$ in terms of a known 
function
but it is still given by the integral in (\ref{soln_f}).  One can 
similarly generalize
this for a collection of $n$ M5(2) branes:
\eq
F=\prod_{i=1}^{n} \sinh(y-y_i)^{2/N}.
\qe
This again determines $f$ in principle through eq.~(\ref{soln_f}).

\section{Conclusions and discussion}
In this paper we presented an exact solution of 11-dimensional 
supergravity describing localized intersections of M5-branes.
The solution has some surprising features worth pointing out.

The geometry of our intersecting brane configuration is a 
warped anti-de Sitter product geometry, consistent with the fact
that the dual quantum field theory is a conformal field theory.

Another feature concerns the `t Hooft coupling.
Taking the large $N$ limit to remain in the domain of validity
of supergravity does not imply anything about the value of
$g_{YM}^2 =R/L$.  This ratio
is an arbitrary constant.  In the solutions known thus far for
4-dimensional field theories the relevant combination appearing
in the supergravity solution is always $g_{YM}^2 N$, 
forcing the `t Hooft coupling
to be large in the small curvature limit relevant to supergravity.  
In our case there appears to be no such restriction on the `t Hooft coupling.
It is thus surprising that in principle one can tune the `t Hooft 
coupling to be small or large while remaining in the domain of 
validity of supergravity. However, the large $N$ limit may be rather subtle in
this case and this issue is currently under investigation.

Our solution does not have any simple $N$ dependence: there are
terms of different orders in $N$ in a $1/N$ expansion 
despite the fact that we have taken the decoupling limit.  
Unlike the $AdS_{5}\times S^5$ case, the $1/N$ suppressed terms do not come
with powers of the Planck scale.  Certainly further terms relevant to
the asymptotically flat solution will contain the Planck scale, 
however, it is surprising that the $1/N$ corrections do not appear 
to be directly connected to an expansion in the Planck scale.

There are a number of directions which open up from this analysis.
One is to consider other intersecting branes which are connected to
this configuration through compactification and T-duality.  
The present system of M5-branes can be viewed as a special case of
a more general problem of an M5-brane wrapped on a Riemann surface.
The supergravity description of the general problem will be of 
interest  
for finding supergravity duals for more interesting field 
theories, including non-conformal field theories.  The solution to 
the latter problem will be presented in \cite{pap2}.

\section{Acknowledgements}
AF would like to thank Subir Mukhopadhyay for discussions, and the
Durham University Department of Mathematical Sciences, where part of this work
was done, for their hospitality. AF is supported by a grant from the Swedish
Research Council, he would also like to acknowledge support from
the NSF under grant PHY99-73935 during the academic year `98-'99 when
this project was initiated.



\begin{thebibliography}{77}
\bibitem{gaunt}
J.~Gauntlett {\it Intersecting Branes} hep-th/9705011.
\bibitem{Itzhaki:1998uz}
N.~Itzhaki, A.A.~Tseytlin and S.~Yankielowicz
{\it Supergravity solutions for branes localized within branes}
Phys. Lett. {\bf B432}, 298 (1998),
hep-th/9803103.
\bibitem{loc2} A. Hashimoto {\it Supergravity solutions for 
localized 
intersections of branes} JHEP {\bf 9901} (1999) 018,
hep-th/9812159.
\bibitem{loc1}H. Yang {\it Localized intersecting brane
solutions of D=11 supergravity} hep-th/9902128.
\bibitem{loc3}D. Youm {\it Localized intersecting BPS branes}
hep-th/9902208.
\bibitem{danda}A. Fayyazuddin and D. J. Smith {\it Localized
intersections of M5-branes and four-dimensional superconformal
field theories} JHEP {\bf 9904} (1999) 030,
hep-th/9902210.
\bibitem{loc3.5}A. Loewy {\it Semi Localized Brane Intersections in 
SUGRA}
Phys. Lett. {\bf B463} {1999} 41, hep-th/9903038.
\bibitem{loc4}A. Gomberoff, D. Kastor, D. Marolf, J. Traschen {\it
Fully Localized Brane Intersections - The Plot Thickens}
Phys. Rev. {\bf D61} (2000) 024012, hep-th/9905094.
\bibitem{loc5}D. Youm {\it Supergravity Solutions for BI Dyons}
Phys. Rev. {\bf D60} (1999) 105006, hep-th/9905155.
\bibitem{loc6}S. A. Cherkis {\it  Supergravity Solution for M5-brane
Intersection} hep-th/9906203.
\bibitem{pema}D. Marolf and A. Peet {\it Brane Baldness vs.
Superselection Sectors} Phys. Rev. {\bf D60} (1999) 105007, hep-th/9903213;
A. W. Peet {\it Baldness/delocalization in intersecting brane
systems} Class. Quant. Grav. {\bf 17} (2000) 1235,
hep-th/9910098.
\bibitem{Maldacena}
J.M.~Maldacena {\it The Large N Limit of Superconformal Field 
Theories and
Supergravity} Adv. Theor. Math. Phys. {\bf 2} (1998) 231-252, hep-th/9711200.
\bibitem{AdS_refs}
S.~Gubser, I.~Klebanov and A.~Polyakov {\it Gauge theory correlators 
from
noncritical string theory} Phys. Lett. {\bf B428} (1998) 105, 
hep-th/9802109;
E.~Witten {\it Anti-de Sitter space and holography} Adv. Theor. Math. 
Phys.
{\bf 2} (1998) 253, hep-th/9802150.
\bibitem{WAdS1}M. Alishahiha and Y. Oz 
{\it AdS/CFT and BPS Strings in Four Dimensions} hep-th/9907206;
Y.~Oz {\it Warped Compactifications and AdS/CFT} hep-th/0004009
\bibitem{WAdS2}
M.~Cveti\v{c}, H.~L\"u, C.N.~Pope, J.F.~V\'azquez-Poritz
{\it AdS in Warped Spacetimes} hep-th/0005246
\bibitem{HW}
A.~Hanany, E.~Witten {\it Type IIB Superstrings, BPS Monopoles, And
Three-Dimensional Gauge Dynamics} Nucl. Phys. {\bf B492} (1997) 
152-190,
hep-th/9611230.
\bibitem{witn2}
E.~Witten {\it Solutions Of Four-Dimensional Gauge Theories Via M 
Theory}
Nucl. Phys. {\bf B500} (1997) 3-42, hep-th/9703166.
\bibitem{vafa} A. Klemm, W. Lerche, P. Mayr, C.Vafa, N. Warner
{\it Self-Dual Strings and N=2 Supersymmetric Field Theory}
Nucl. Phys. {\bf B477} (1996) 746, hep-th/9604034.
\bibitem{malstrings}J.M.~Maldacena
{\it Branes probing black holes} Nucl. Phys. {\bf B513} (1998) 198,
hep-th/9708147.
\bibitem{pap2}B. Brinne,
A.~Fayyazuddin, S. Mukhopadhyay and D.J.~Smith {\it In preparation}.


\end{thebibliography}
\end{document}